\documentclass[twocolumn,showpacs,preprintnumbers,amsmath,amssymb]{revtex4-1}

\usepackage{graphicx}
\usepackage{dcolumn}
\usepackage{bm}
\usepackage{color}

\begin{document}

\title{Optically-gated resonant emission in single quantum dots}
\author{H. S. Nguyen$^{1}$, G. Sallen$^{1,2}$, C. Voisin$^{1}$, Ph. Roussignol$^{1}$, C. Diederichs$^1$ and G. Cassabois$^{1,3,4}$}
\affiliation{$^{1}$Laboratoire Pierre Aigrain, Ecole Normale Sup\'{e}rieure, CNRS (UMR 8551),
Universit\'{e} P. et M. Curie, Universit\'{e} D. Diderot, 24, rue Lhomond, 75231 Paris Cedex 05, France\\
$^2$Laboratoire de Physique et Chimie des Nano-Objets, 135 avenue de Rangueil, 31077 Toulouse, France\\
$^3$Universit\'e Montpellier 2, Laboratoire Charles Coulomb UMR5221, F-34095, Montpellier, France\\
$^4$CNRS, Laboratoire Charles Coulomb UMR5221, F-34095, Montpellier, France}
\date{\today}

\begin{abstract}
We report on the resonant emission in coherently-driven single semiconductor quantum dots. We demonstrate that an ultra-weak non-resonant laser acts as an optical gate for the quantum dot resonant response. We show that the gate laser suppresses Coulomb blockade at the origin of a resonant emission quenching, and that the optically-gated quantum dots systematically behave as ideal two-level systems in both regimes of coherent and incoherent resonant emission.
\end{abstract}

\maketitle

Semiconductor quantum dots (QDs) are commonly considered as artificial atoms \cite{qdot}, and this analogy has led to an important development of experiments in the field of cavity quantum electrodynamics with semiconductor systems. These nanostructures are very promising for the realization of integrated devices such as single photon sources for quantum cryptography applications. The corresponding photon antibunching effect was observed by photoluminescence experiments in single QDs under non-resonant excitation \cite{Michler:00}, and more recently under strictly resonant excitation of the fundamental interband transition \cite{Muller:07,Ates:09}. In particular, the application of this technique to single QDs in the weak coupling regime has led to the generation of indistinguishable photons thanks to the shortening of the radiative lifetime due to the Purcell effect, and to the reduction of the dephasing processes due to the strictly resonant excitation of the QD \cite{Ates:09}. In fact, this configuration minimizes the residual excitation of the QD environment that acts as a fluctuating reservoir for spectral diffusion \cite{Kammerer:02,Berthelot:06}. In this context, the intrinsic properties of a two-level system can be studied such as the Rabi oscillations \cite{Muller:07,Melet:08,Flagg:09} or the Mollow triplet in the resonant fluorescence regime \cite{Flagg:09}.

In this letter, we show that experimental studies of the resonant emission (RE) of single QDs can be strongly limited, even impossible, since most of the QDs show a very strong quenching of the RE response of the neutral exciton. In the prospect of efficient generation of polarization-entangled Bell states in integrated devices \cite{Fattal:04,dousse:10}, this effect is highly problematic. We propose an efficient way to overcome this issue based on the use of an ultra-weak non-resonant laser that optically gates the QD RE response. The optical gate induces a complete recovery of the properties of an artificial atom, and the single QDs behave as ideal two-level systems. We show that the QD RE is in fact quenched by Coulomb blockade, and that the non-resonant gate laser controls the QD ground state, even in the coherent regime of Rayleigh scattering.

Resonant excitation experiments on single QDs are performed at low temperatures at 10~K with an orthogonal excitation-detection setup \cite{Muller:07,Melet:08}. The sample consists of a low density layer (0.01~$\mu m^{-2}$) of self-assembled InAs/GaAs QDs embedded in a planar $\lambda_0$-GaAs microcavity formed by two distributed-Bragg reflectors. The top and bottom Bragg mirrors have 11 and 24 pairs of $\lambda_0/4$ layers of AlAs and AlGaAs, respectively. The QD emission is detected in the microcavity mode, which has a quality factor $Q$ of 2500 and a central energy of 1.27 eV, by means of a conventional confocal setup. The excitation laser is injected into the in-plane guided mode supported by the planar microcavity. The QD resonant excitation is performed either by a tunable cw external cavity diode laser for experiments in the spectral domain with an overall resolution of 1 $\mu$eV, or by a cw mode-locked Ti:Sa laser for time domain measurements with a temporal resolution of 90 ps. Fourier transform spectroscopy enables high spectral resolution measurements of the QD emission \cite{Berthelot:06}. The non-classical properties of the QD resonant emission is further investigated by a Hanbury-Brown and Twiss (HBT) setup for intensity correlation measurements \cite{Michler:00}. Finally, the optical gate at the heart of our study is provided by a cw HeNe laser at 632~nm (1.9 eV) that is vertically injected into the planar microcavity via the confocal setup.

\begin{figure}[htb]
\begin{center}
\includegraphics[width=8.5cm]{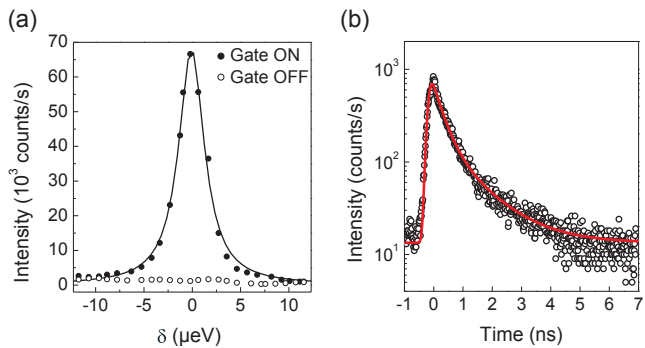}
\end{center}
\caption{\label{GatedRE}{(a) Resonant emission (RE) spectra of the exciton with ($\bullet$, gate ON) and without ($\circ$, gate OFF) the additional non-resonant laser, in a single QD at 10K. The RE intensity is measured as a function of the laser detuning with the exciton energy, $\delta=E_{L}-E_{X}$. The powers of the optical gate and the resonant laser are respectively equal to 3~nW and $4~\mu W$. Lorentzian fit (solid line) of width $3.2~\mu eV$. (b) Time-resolved RE of the QD when the gate is switched on ($\circ$). The solid line is a fit with a decay time $T_1=330\pm 20$~ps.}}
\end{figure}

The implementation of the resonant excitation technique appears to be successful for very few QDs. The reproduction of the typical results reported by Muller \textit{et al.} in Ref.\cite{Muller:07} is achieved for 1 QD over 50, and the most frequent case corresponds to no resonance in the RE spectrum of the neutral exciton (Fig.\ref{GatedRE}(a), open circles $\circ$). Since the excitation laser is strictly at the energy of the QD fundamental interband transition, our observation strongly contradicts the expectations for having a quasi-ideal two-level system, which are based on the upmost reduction of the environment fluctuations under resonant excitation of the QD \cite{Kammerer:02,Bayer:02,Langbein:04,Hogele:04}. We demonstrate below a solution with an optical gate.

We optically control the QD resonant response by means of a non-resonant laser that systematically restores the RE response of the neutral exciton in the studied QDs. Figure~\ref{GatedRE}(a) shows the RE spectrum of a single QD when the optical gate is switched on ($\bullet$) and off ($\circ$), and we stress that this result corresponds to the general behavior. The intensity of the RE is measured as a function of the laser detuning $\delta=E_{L}-E_{X}$, where $E_L$ and $E_X$ are the laser and the exciton energies. While almost all the QDs show no resonance when the optical gate is switched off, a clear resonance appears at zero detuning when the optical gate is switched on. The optically-gated RE spectrum is fitted with a Lorentzian line of width 3.2~$\mu$eV, comparable to the typical linewidths measured under resonant excitation with various techniques \cite{Hogele:04,Langbein:04,Muller:07}. We stress that the power of the additional non-resonant laser is typically of few nanowatts, that is more than four decades lower than the power required to saturate the QD. The contribution of the photoluminescence that is non-resonantly excited by the optical gate is therefore completely negligible.

In the following, we present a complete experimental study on the optically-gated RE in single QDs. We show that the intrinsic properties of a two-level system under resonant excitation are recovered when the optical gate is switched on. Figure~\ref{GatedRE}(b) first presents time-resolved measurements of the optically-gated QD emission under resonant excitation ($\circ$). A typical exponential decay is observed in the temporal trace and a fit (solid line) gives the QD lifetime $T_1=330\pm 20$~ps. This estimate is a standard value for epitaxially grown semiconductor QDs \cite{qdot}.

\begin{figure}[htb]
\begin{center}
\includegraphics[width=8.5cm]{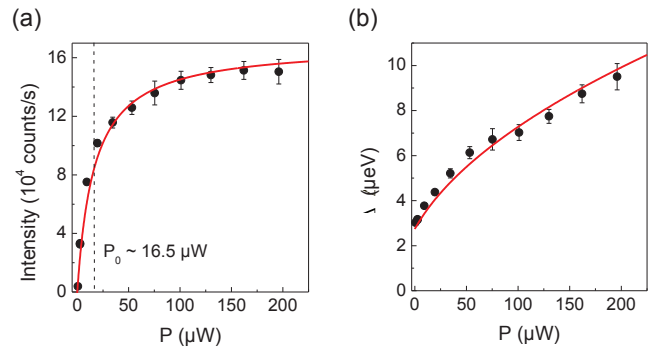}
\end{center}
\caption{\label{2LevelSystem}{Intensity (a) and linewidth (b) of the optically-gated resonant emission spectrum versus the power of the resonant laser, P. The experimental data are fitted by using Eq.1 and 2 corresponding to a two-level system model (solid line).}}
\end{figure}

We now turn to the power-dependence of the optically-gated RE, which allows for the evaluation of the decoherence time $T_2$. Figure~\ref{2LevelSystem} displays the intensity at zero detuning ($\delta=0$) (Fig.~\ref{2LevelSystem}(a)) and the linewidth $\Delta$ (Fig.~\ref{2LevelSystem}(b)) of the optically-gated RE spectrum as a function of the power of the resonant diode laser. The intensity of the RE signal first increases linearly with the excitation power, and then displays the typical saturation expected at high powers. In parallel, we observe the text-book power-broadening of the RE spectrum, with a linewidth increasing by a factor of three in our investigated range of excitation power. The theoretical expressions of the emission intensity $I$ at zero detuning ($\delta=0$) and the linewidth $\Delta$ of a two-level system under resonant excitation are \cite{AtomePhoton}:
\begin{eqnarray}
I\propto\frac{\Omega^2T_1T_2}{1+\Omega^2T_1T_2}&\propto&\frac{P}{P+P_0}\\
\Delta=\frac{2\hbar}{T_2}\sqrt{1+\Omega^2T_1T_2}&=&\frac{2\hbar}{T_2}\sqrt{1+\frac{P}{P_0}}
\end{eqnarray}
where $\Omega$ is the Rabi frequency, $T_1$ the population lifetime, $T_2$ the decoherence time, $P$ the power of the resonant laser and $P_0$ the saturation power. The experimental data presented in Fig.~\ref{2LevelSystem} are perfectly well fitted to equations (1) and (2), with $T_2=480\pm10$~ps and $P_0=16.5\pm1.5~\mu$W. We then deduce from the extracted value of $T_2$ that $T_2\sim 1.5T_1$, which means the QD homogeneous linewidth is very close to its radiative limit ($T_2=2T_1$). More precisely, our experimental data indicate that the pure dephasing time $T_2^\star$ is of the order of 1.8~ns, $T_2^\star$ being defined by $1/T_2=1/2T_1+1/T_2^\star$. Such a pure dephasing time attests for the reduction of the environment fluctuations under resonant excitation where spectral diffusion effects are minimized \cite{Kammerer:02,Berthelot:06}. We further conclude that the intrinsic features of a two-level system under resonant excitation are clearly restored by the ultra-weak optical gate. In particular, our technique is validated in both regimes of coherent and incoherent RE. For $P\ll P_0$, it is known that the RE consists in the coherent signal of Rayleigh scattering, whereas at saturation, the incoherent photoluminescence dominates the RE \cite{AtomePhoton}. In figure~\ref{GatedRE}(a), for a given detuning, the recorded intensity consists in the spectrally integrated signal of both components.
\begin{figure}[htb]
\begin{center}
\includegraphics[width=8.5cm]{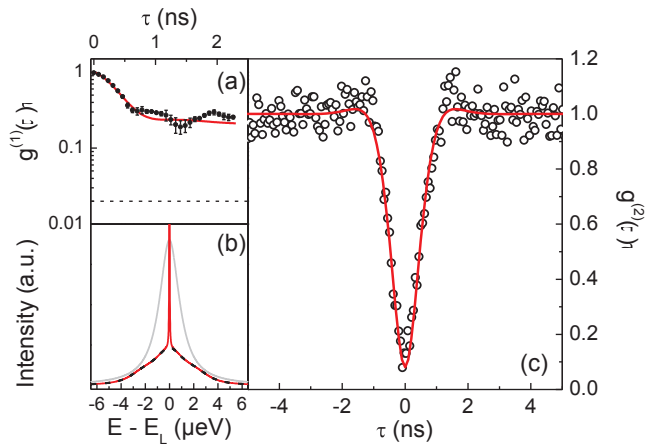}
\end{center}
\caption{\label{AB}{(a) High-resolution Fourier transform spectroscopy of the optically-gated RE, at zero detuning ($\delta=0$), for $P/P_0=2.4$. The background level is indicated with a dashed line. Fit (solid line) by using Eq.~3 with $T_L=100$~ns, $\hbar\Omega=2.5~\mu$eV, $T_1=330$~ps and $T_2=480$~ps. (b) Theoretical RE signal in the spectral domain (red solid line) together with the incoherent contribution (dashed line) and the radiatively-limited emission line of width 2~$\mu$eV (gray solid line). (c) Intensity auto-correlation measurements under resonant excitation at $\delta=0$, for $P/P_0=0.7$, when the optical gate is switched on.}}
\end{figure}
High-resolution measurements by Fourier transform spectroscopy of the RE at zero detuning ($\delta=0$) for $P/P_0=2.4$ show the superposition of the coherent and incoherent components of the RE (Fig.~\ref{AB}(a \& b)). Our data are well fitted (solid lines in Fig.~\ref{AB}(a \& b)) to the first order correlation function $g^{(1)}(\tau)$ of the RE of a two-level system \cite{sm}:
\begin{eqnarray}
g^{(1)}(\tau)&\propto&\text{e}^{-\frac{\tau}{T_L}} + A\left[\text{e}^{-\frac{\tau}{T_2}}+\text{e}^{-\alpha\tau}\left(B\cos(\beta\tau+\phi)\right)\right]
\end{eqnarray}
where $T_L$ is the coherence time of the resonant laser, $A$, $B$ and $\phi$ are three constants which depend on $\Omega$, $T_1$ and $T_2$, $\alpha=\frac{1}{2}\left(\frac{1}{T_{1}}+\frac{1}{T_{2}}\right)$, and $\beta=\sqrt{\Omega^2-\frac{1}{4}\left(\frac{1}{T_{1}}-\frac{1}{T_{2}}\right)^2}$. The first term in the $g^{(1)}(\tau)$ function corresponds to the RE signal that is coherent with the excitation laser, whereas the second one is related to the incoherent part. In Fig.~\ref{AB}(b), we display the Fourier transform of the theoretical $g^{(1)}(\tau)$ function which gives the RE signal in the spectral domain. The incoherent component (dashed line in Fig.~\ref{AB}(b)) is broader than the radiatively-limited emission line (gray line in Fig.~\ref{AB}(b)) of 2~$\mu$eV width. On the contrary, the central peak of the coherent resonant Rayleigh scattering is much narrower than the natural line. Indeed, we demonstrate that the ultra-weak non-resonant optical gate induces the recovery of the properties of an artificial atom even in the coherent regime of Rayleigh scattering.

We finally discuss the quantum properties of the optically-gated RE signal in single QDs. Figure~\ref{AB}(c) presents the normalized intensity auto-correlation measurements by the HBT setup under resonant excitation at zero detuning ($\delta=0$), when the optical gate is switched on. After subtracting the noise contribution, we observe a pronounced dip at zero time delay with $g^{(2)}(0)=0.06$. This is a clear evidence of the characteristic photon antibunching in a single photon source \cite{Michler:00,Muller:07,Ates:09}. Most importantly, it further shows that the ultra-weak gate laser does not degrade the non-classical properties of a single QD emission. Our data are well fitted by the convolution of the system response function of 150~ps resolution with the theoretical intensity auto-correlation function $g^{(2)}(\tau)$ of the RE of a two-level system, given by $g^{(2)}(\tau)=1-\left(\cos(\beta\tau)+\frac{\alpha}{\beta}\sin(\beta\tau)\right)\text{e}^{-\alpha\tau}$ \cite{QO,Flagg:09}. Taking into account the previous measured values of $T_1$ and $T_2$, the data are fitted with $\hbar\Omega=1.4\pm0.1~\mu eV$. The QD dipole moment, defined as $\mu=\hbar\Omega/E$ with $E$ the electric field amplitude, is then deduced from the fitted value of the Rabi frequency. From the power of the resonant excitation laser $P=11.5~\mu$W (measured before coupling into the optical setup), we deduce $E\sim0.013$~kV/cm so that we obtain an estimation of $\mu\sim50$~D. This dipole moment is in agreement with other estimations in similar InAs QDs ranging from 20 to 60~D \cite{Muller:04,Yoshie:04,Borri:02,Kamada:01}.

\begin{figure}[htb]
\begin{center}
\includegraphics[width=8.5cm]{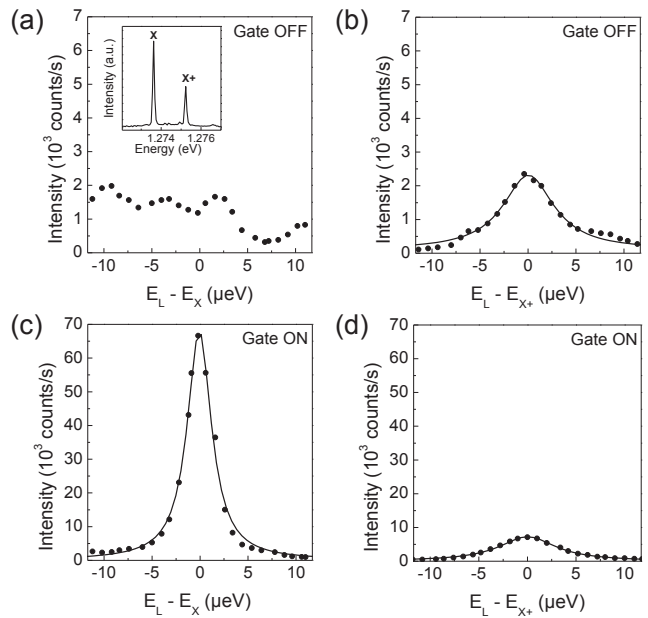}
\end{center}
\caption{\label{X-CX}{RE intensity as a function of the energy detuning between the resonant laser and the neutral exciton X (a \& c), and positively charged exciton $\text{X}^+$ (b \& d), when the gate laser is switched off (a \& b) and on (c \& d). The solid lines are Lorentzian fits of FWHM $3.2~\mu$eV (c) for the neutral exciton, and $6.5~\mu$eV (b) and $7.1~\mu$eV (d) for the charged one. Inset: micro-photoluminescence spectrum under non-resonant excitation showing the neutral and charged excitons.}}
\end{figure}

We interpret the optical gate-effect within a model where the QD RE is quenched by Coulomb blockade because of the tunneling of a hole from a structural deep level \cite{sm}. The optical gate suppresses Coulomb blockade by switching back the QD into a neutral state. This interpretation is supported by additional measurements of the QD RE, for the two spectral windows of the neutral exciton (X) and positively charged exciton ($\text{X}^+$), centered at $E_X$ and $E_{\text{X}^+}$. The inset of Fig.~\ref{X-CX}(a) displays a micro-photoluminescence spectrum under non-resonant excitation showing X and $\text{X}^+$ (see \cite{sm} for the identification of the two lines). When the optical gate is switched off, no resonance is observed for the neutral exciton, as displayed in Fig.~\ref{X-CX}(a) (and already shown in Fig.~\ref{GatedRE}(a)), whereas RE is observed for the charged exciton (Fig.~\ref{X-CX}(b)) (the higher background signal for X (Fig.~\ref{X-CX}(a) and (b)) comes from the different detunings of the neutral and charged emission lines with the cavity mode, the X line being in resonance with the cavity mode in the present case). On the contrary, when the optical gate is switched on, the RE of the neutral exciton (Fig.~\ref{X-CX}(c)) appears and becomes more intense than for the charged one (Fig.~\ref{X-CX}(d)). These experimental data indicate that the optical gate controls the residual charge $q$ in the QD. When the optical gate is switched off, the $q$ value is close to +e whereas $q$ becomes closer to 0 when the optical gate is switched on.

The electrical space-charge technique of deep level transient spectroscopy brought evidence that the growth process of self-assembled QDs results in the intrinsic coexistence of structural deep levels and QDs \cite{Lin:05}. In fact, the photophysics of defects in arsenide semiconductors presents a rich phenomenology. For instance, in fractional quantum Hall effect experiments, the mobility of two-dimensional electron systems is increased to the highest values by using a weak non-resonant laser \cite{eisenstein:88}. In bulk semi-insulating GaAs samples, persistent photoconductivity arises from the metastability of the arsenic antisite, due to the complex bond-breaking structural rearrangement in GaAs \cite{Vincent:82,Chadi:88,dabrowski:88}. In our case, the structural defects around the QD contribute to the presence of residual charges inside the QD, which lead to the coexistence of neutral and charged exciton lines (Fig.~\ref{X-CX}(a), inset). Such a feature is routinely observed in QD samples from different growth facilities \cite{finley:02,Moskalenko:03,igarashi:10} or made of various materials \cite{Hartmann:00,Besombes:02}. Several papers reported the optical control of the QD residual charge in photoluminescence spectroscopy under non-resonant excitation \cite{Hartmann:00,Moskalenko:03}. Although analog to our optical-gate effect, it strongly contrasts with our experiments. The techniques used in Ref.\cite{Hartmann:00,Moskalenko:03} refer to a complete incoherent regime where the photoluminescence is non-resonantly excited. Our work demonstrates a novel approach for coherently-driven single QDs. In particular, even in the regime of coherent Rayleigh scattering, where no energy is absorbed in the QD \cite{Mukamel}, the non-resonant gate laser finely controls the QD ground state. This configuration opens the way for conditional quantum measurements time-controlled by the gate laser.

In summary, we have shown that experimental studies of the resonant emission of coherently-driven single QDs can be strongly limited, even impossible, since most of the QDs show no resonance in the resonant emission spectrum of the neutral exciton. We demonstrate a solution based on an ultra-weak non-resonant laser that optically gates the QD resonant response. The optical gate induces a complete recovery of the properties of an artificial atom, and the single QDs behave as ideal two-level systems. We show that the QD resonant emission is in fact quenched by Coulomb blockade, and that the non-resonant gate laser controls the QD ground state, even in the coherent regime of Rayleigh scattering.

The authors gratefully acknowledge P. M. Petroff for the sample fabrication, as well as G. Bastard, R. Ferreira and C. Delalande for stimulating discussions. The work was supported financially by the 'Agence Nationale pour la Recherche' project QSWITCH.\\

\end{document}


\title{Supplemental material for: Optically-gated resonant emission in single quantum dots}
\author{H. S. Nguyen$^{1}$, G. Sallen$^{1,2}$, C. Voisin$^{1}$, Ph. Roussignol$^{1}$, C. Diederichs$^1$ and G. Cassabois$^{1,3,4}$}
\affiliation{$^{1}$Laboratoire Pierre Aigrain, Ecole Normale Sup\'{e}rieure, CNRS (UMR 8551),
Universit\'{e} P. et M. Curie, Universit\'{e} D. Diderot, 24, rue Lhomond, 75231 Paris Cedex 05, France\\
$^2$Laboratoire de Physique et Chimie des Nano-Objets, 135 avenue de Rangueil, 31077 Toulouse, France\\
$^3$Universit\'e Montpellier 2, Laboratoire Charles Coulomb UMR5221, F-34095, Montpellier, France\\
$^4$CNRS, Laboratoire Charles Coulomb UMR5221, F-34095, Montpellier, France}

\maketitle

\textbf{I. Coherent and incoherent contributions in the resonant emission.}

Figure~3(a \& b) display high-resolution measurements by Fourier transform spectroscopy of the resonant emission (RE). The coherent and incoherent components of the RE are fitted by the first order correlation function $g^{(1)}(\tau)$ of the RE of a two-level system:
\begin{eqnarray*}
g^{(1)}(\tau)&\propto&\text{e}^{-\frac{\tau}{T_L}} + A\left[\text{e}^{-\frac{\tau}{T_2}}+\text{e}^{-\alpha\tau}\left(B'\cos(\beta\tau)+C'\sin(\beta\tau)\right)\right]
\end{eqnarray*}
where
\begin{equation*}
	\begin{aligned}
	  \alpha &=\frac{1}{2}\left(\frac{1}{T_{1}}+\frac{1}{T_{2}}\right)\\
		\beta &=\sqrt{\Omega^2-\frac{1}{4}\left(\frac{1}{T_{1}}-\frac{1}{T_{2}}\right)^2}\\
        A &=\frac{T_1}{T_2}(1+\Omega^2T_1T_2) \\
		B' &=1-\frac{T_2}{T_1(1+\Omega^2T_1T_2)} \\
	  C' &=\frac{\Omega^2T_1(3T_2-T_1)-\frac{(T_1-T_2)^2}{T_1T_2}}{2\beta T_1(1+\Omega^2T_1T_2)}
	\end{aligned}
\end{equation*}

The Fourier transform of the $g^{(1)}(\tau)$ function gives the RE signal in the spectral domain $I(\omega)$, which is the sum of the coherent and incoherent components, $I_\text{coh}(\omega)$ and $I_\text{incoh}(\omega)$ respectively, in the RE spectrum:

\begin{equation*}
    \begin{aligned}
        I_\text{coh}(\omega) &\propto \frac{\Omega^2T_2^2}{(1+\Omega^2T_1T_2)^2}\frac{{\frac{2}{T_L}}}{(\omega-\omega_L)^2+\frac{1}{T_L^2}}\\
        I_\text{incoh}(\omega)&\propto \frac{\Omega^2T_1T_2}{1+\Omega^2T_1T_2}[\frac{{\frac{2}{T_2}}}{(\omega-\omega_0)^2+\frac{1}{T_2^2}}\\
        &+ \frac{\alpha B'+(\omega+\beta-\omega_0)C'}{(\omega+\beta-\omega_0)^2+\alpha^2}\\
        &+ \frac{\alpha B'-(\omega-\beta-\omega_0)C'}{(\omega-\beta-\omega_0)^2+\alpha^2}]
    \end{aligned}
\end{equation*}

with $E_L=\hbar\omega_L$ the laser energy and $\hbar\omega_0$ the energy of the resonantly-excited fundamental transition. In particular, $I_\text{coh}(\omega)$ corresponds to the central peak of the coherent resonant Rayleigh scattering observed in Fig.~3(b). Moreover, in the configuration where $P\sim P_0$, the incoherent part of the RE signal is limited to the first term of $I_\text{incoh}(\omega)$.\\

\textbf{II. Identification of the neutral and charged excitons.}

The inset of figure~4(a) displays a typical micro-photoluminescence spectrum under non-resonant excitation showing two lines. The lower and upper lines are labelled X and CX, respectively. We show below by auto- and cross-correlation Hanbury-Brown and Twiss (HBT) measurements (Fig.~\ref{info1}) that the X line corresponds to the neutral exciton, and the CX one to the charged exciton. The $g^{(2)}$ correlation functions are displayed in Fig.~\ref{info1} after deconvolution by the system response function.

\renewcommand{\thefigure}{S\ \arabic{figure}}
\begin{figure}[htb]
\begin{center}
\includegraphics[width=8.5cm]{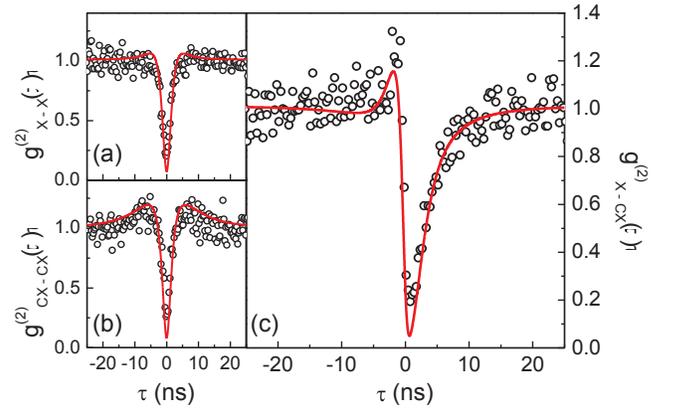}
\end{center}
\caption{{\label{info1}}Intensity auto- (a \& b) and cross- (c) correlation measurements of the X and CX lines observed in the inset of figure~4(a). The experimental data are fitted by using the rate equations corresponding to the model in Ref.~\cite{Gomis:11} (solid lines). The X and CX lines are identified as the neutral and charged excitons, respectively.}
\end{figure}

We observe a pronounced dip at zero time delay for all three measurements, with $g^{(2)}_{X-X}(0)=0.08$ and $g^{(2)}_{CX-CX}(0)=0.09$ for the auto-correlation functions of X and CX, respectively, and $g^{(2)}_{X-CX}(0)=0.05$ for the cross-correlation function between the two lines. The anti-bunching in the cross-correlation experiment (Fig.~\ref{info1}(c)) is a clear evidence that the two lines observed in the micro-photoluminescence spectrum correspond to the same quantum dot (QD). Moreover, since $g^{(2)}_{X-CX}$ corresponds to the probability of detecting one CX photon after one X photon for positive time delays, the bunching at negative time delays allows us to identify the CX line as corresponding to a charged state of the X line \cite{Gomis:11}. All the results are well fitted by using the rate equations corresponding to the model in Ref.~\cite{Gomis:11}. In a simple manner, the long time part of $g^{(2)}(\tau >0)$ is attributed to a three charges capture process whereas the short time part of $g^{(2)}(\tau <0)$ is attributed to a single charge capture process. Finally, from the the positive detuning of the charged exciton state (CX) with  respect to the neutral one (X), the CX line is attributed to the positive trion made of of two holes and one electron \cite{Regelman:01,Ediger:05}.\\

\textbf{III. Model of the optical gate-effect}

We interpret the optical gate-effect within a model where the QD resonant emission (RE) is quenched by Coulomb blockade because of the tunneling of a hole from a structural deep level (DL). The optical gate suppresses Coulomb blockade by switching back the QD into a neutral state.

\begin{figure}[htb]
\begin{center}
\includegraphics[width=8.5cm]{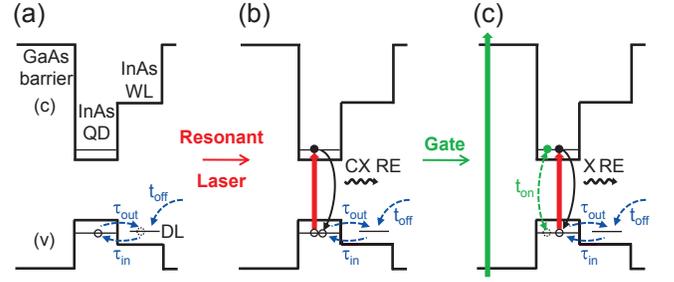}
\end{center}
\caption{{\label{info2}}Scheme of the valence (v) and conduction (c) band-edges in the growth direction of the heterostructure, showing together the InAs wetting layer (WL), the GaAs barrier, the single QD and a deep level (DL) assumed to be close to the confined hole state of the QD, with electrons ($\bullet$) and holes ($\circ$). (a) The resonant laser and the optical gate are switched off. A hole tunnels from the DL to the QD and vice-versa, with characteristic times $\tau_{\text{in}}$ and $\tau_{\text{out}}$, respectively. The DL filling with a rate 1/$t_{\text{off}}$ leads to a residual hole in the QD. (b) The resonant laser (shown by the up solid arrow) is switched on. The QD resonant emission (RE) is observed only at the energy $E_{CX}$ of the positively charged exciton CX. (c) The resonant laser and the optical gate (long solid arrow) are switched on. The residual hole is ejected from the QD after radiative recombination with one electron photogenerated by the optical gate, with a time constant equal to $t_{\text{on}}$. The DL recaptures a hole with a time constant equal to $t_{\text{off}}$. $t_{\text{on}}<t_{\text{off}}$ leads to a neutral QD state, and the QD RE is predominantly observed at the energy $E_X$ of the neutral exciton X.}
\end{figure}

The processes governing the optical gate-effect are represented in the model sketched in figure \ref{info2}. The defect state is assumed to be close to the QD confined hole state. When the resonant laser and the optical gate are switched off (Fig.~\ref{info2}(a)), the hole tunnels from the structural deep level to the QD with a time constant $\tau_{in}$, and from the QD back to the deep level with a time constant $\tau_{out}$, the deep level being filled with a rate $1/t_{\text{off}}$. This leads to a residual charge $q$ close to +e in the QD. When the QD is resonantly excited by the resonant laser (Fig.~\ref{info2}(b)), only the RE of CX is observed since the residual hole in the QD blocks any optical transition at the energy $E_X$ of the neutral exciton because of Coulomb blockade \cite{Hogele:04}.

When the optical gate is switched on (Fig.~\ref{info2}(c)), the residual hole is ejected from the QD through radiative recombination with one electron photogenerated by the optical gate in the GaAs barrier. The inverse of the $t_{\text{on}}$ time thus characterizes the corresponding rate of radiative recombination. This process is extremely rare since the optical gate is more than four orders of magnitude lower than at the QD saturation. However, when it occurs, it switches the QD back into a neutral state where the RE of the neutral exciton is observable (Fig.~1(a) and 4(c)). The probability of having no residual hole in the QD is then limited by the recharging of the deep level with a time constant $t_{\text{off}}$.

This phenomenology is well described by a simple rate equations model which gives a rough estimation of the transient time constants $t_{\text{on}}$ and $t_{\text{off}}$ measured with time-resolved experiments (these results on the switching dynamics of the optical gate effect will be the subject of a forthcoming publication). The corresponding rate equations are:
\begin{eqnarray*}
\frac{dp_{DL}}{dt}&=&-\frac{p_{DL}(1-p_{QD})}{\tau_{in}}+\frac{p_{QD}(1-p_{DL})}{\tau_{out}}+G(1-p_{DL})\\
\frac{dp_{QD}}{dt}&=&+\frac{p_{DL}(1-p_{QD})}{\tau_{in}}-\frac{p_{QD}(1-p_{DL})}{\tau_{out}}-\mu P_Gp_{QD}
\end{eqnarray*}
where $p_{QD}$ and $p_{DL}$ are the probabilities to have one hole in the QD and DL, respectively, $G$ the capture rate of one hole in the DL (assumed to be constant at low optical gate powers), and $\mu P_G$ the radiative recombination rate of the residual hole in the QD, $P_G$ being the optical gate power. This model leads to a transient time $t_{\text{on}}$ of the order of $1/\mu P_G$, and to a $t_{\text{off}}$ time of the order of $1/G$. In the limit of the QD saturation by the optical gate, $\mu P_G$ corresponds to the neutral exciton recombination rate $\gamma_R$, with $1/\gamma_R=T_1\sim500$~ps. Therefore, we rewrite $\mu P_G$ as $\gamma_R\frac{P_G}{P_G^{sat}}$ with $P_G^{sat}$ the QD saturation power of few tens of $\mu$W. In the experimental conditions where $P_G\sim$ few nW, we get a rough estimate of $t_{\text{on}}$ of 5~$\mu$s, fully consistent with the measured one (not shown here). Moreover, measurements on increasing the optical gate power (not shown here) show (i) a decrease of $t_{\text{on}}$ as a function of $P_G$, and (ii) an approximately constant $t_{\text{off}}$ value, as predicted by our model. Finally, the higher the ratio $t_{\text{off}}/t_{\text{on}}$, the lower the residual charge $q$ in the QD and the more intense the RE of the neutral exciton.

Eventually, we discuss other additional measurements as a function of the energy of the optical gate (not shown here). Remarkably, the intensity of the QD RE varies with the optical gate energy exactly as the QD photoluminescence non-resonantly excited by the intense optical gate. This similarity of the excitation spectra of the RE and non-resonant photoluminescence confirms the assumption of radiative recombination as the process involved in the neutralization of the QD by the optical gate.